\documentclass[a4paper,journall]{IEEEtran}

\usepackage[utf8]{inputenc}
\usepackage{textcomp}
\usepackage{url}
\usepackage{graphicx}
\usepackage{subcaption}
\usepackage[english]{babel}
\usepackage{amsmath}
\usepackage{amssymb}
\usepackage{csvsimple}
\usepackage{multirow}
\usepackage{array}
\usepackage[version=4]{mhchem}
\usepackage{longtable,tabularx}
\usepackage[table]{xcolor}
\usepackage{wrapfig}
\usepackage{tikz}
\usetikzlibrary{positioning,shapes}
\usetikzlibrary{arrows}
\usetikzlibrary{graphs}
\usetikzlibrary{automata}
\tikzset{node distance=2cm, 
every state/.style={ 
semithick,
fill=gray!10},
initial text={}, 
double distance=2pt, 
every edge/.style={ 
draw,
->,>=stealth', 
auto,
semithick}}

\graphicspath{{figures/}{figures/newmaps/}{figures/aida/}}

\newcommand{\nth}{$^\text{th}$}

\newcommand{\nd}{$^\text{nd}$}
\newcommand{\st}{$^\text{st}$}

\title{Analyzing and comparing door-to-door travel times for air transportation using aggregated Uber data}
\author{\IEEEauthorblockN{Philippe Monmousseau, Aude Marzuoli, Eric Feron and Daniel Delahaye}}
\begin{document}
\maketitle
\begin{abstract}
%
Improving the passenger air travel experience is one of the explicit goals set by the Next Generation Air Transportation System in the United States and by the Advisory Council for Aeronautics Research in Europe FlightPath 2050. Both suggest door-to-door travel times as a potential metric for these objectives. 
In this paper, we propose a data-driven model to estimate door-to-door travel times and compare the reach and performance of different access modes to a city, as well as conduct segment analysis of full door-to-door trips.
This model can also be used to compare cities with respect to the integration of their airport within their road structure. We showcase multiple applications of this full door-to-door travel time model to demonstrate how the model can be used to locate where progress can be made.

%
\end{abstract}

%

\section{Introduction}
\label{sec:intro}

Both in Europe and in the United States, national or supra-national agencies promote the need for seamless door-to-door travel and data sharing. They were deemed as needed by the European Commission's 2011 White Paper \cite{darecki2011Flightpath2050Europe} and were reconfirmed by the Federal Aviation Administration (FAA) in 2017 \cite{2017NextGenPrioritiesJoint}. Data sharing was already a main focus in the early 2000s; in response, Europe created and adopted SWIM - System Wide Information Management \cite{meserole2006WhatSystemWide} - and the FAA followed suit. The Next Generation Air Transportation System (NextGen) \cite{nextgen} in the United States and the Advisory Council for Aeronautics Research in Europe (ACARE) Flightpath 2050 \cite{darecki2011Flightpath2050Europe} both aim to have a more passenger-centric approach. To this end, ACARE Flightpath 2050 sets some ambitious goals, which are not all measurable yet due to lack of available data. For example, it aims at having 90\% of travelers within Europe being able to complete their door-to-door journey within 4 hours. In the US, the Joint Planning and Development Office has proposed and tested metrics regarding NextGen's goals \cite{gawdiak2011NextGenMetricsJoint}, but the passenger-centric metrics, especially regarding door-to-door travel times, are still missing. 

Cook et al. \cite{cook2012PassengerOrientedEnhancedMetrics} first explored the shift from flight-centric to passenger-centric metrics in the project POEM - Passenger Oriented Enhanced Metrics - where they propose propagation-centric and passenger-oriented performance metrics and compare them with existing flight-centric metrics. Later, Laplace et al. \cite{laplace2014METACDMMultimodalEfficient} introduce the concept of Multimodal, Efficient Transportation in Airports and Collaborative Decision Making (META-CDM); they propose to link both airside CDM and landside CDM, thus taking passenger perspective into account. In this perspective, Kim et al. \cite{kim2013AirportGateScheduling} propose an airport gate scheduling model for improved efficiency with a balance between aircraft, operator and passenger objectives. Dray et al. \cite{dray2015AirTransportationMultimodal} illustrate the importance of multimodality by considering ground transportation as well during major disturbances of the air transportation system in order to offer better solutions to passengers.

The estimation of door-to-door travel time for multi-modal trips has been previously studied, but for trips contained within the same metropolitan area. Peer et al. \cite{peer2013DoortodoorTravelTimesa} focus on commutes within a Dutch city by studying door-to-door travel times and schedule delays for daily commuters, and show that, for the estimation of the overall travel time, it is important to consider the correlation of travel times across different road links. Salonen and Toivonen \cite{salonen2013ModellingTravelTime} investigate the need for comparable models and measures for trips by car or public transport with focus on the city of Helsinki. Their multi-modal approach takes into account the walking and waiting times necessary to reach a station or a parking place. Duran-Hormazabal and Tirachini \cite{duran-hormazabal2016EstimationTravelTime} analyze travel time variability for multi-modal trips within the city of Santiago, Chile, using human surveyors and GPS data to estimate the time spent in the different transportation modes, namely walking, driving a car, riding a bus and taking the subway. Pels et al. \cite{pels2003AccessCompetitionAirports} analyze the relative importance of access time to airports in the passengers' choice within the San Francisco Bay Area based on a passenger survey, offering perspective from air transportation. These works emphasize the importance of considering all relevant modes when estimating door-to-door travel times, but are limited in scope with respect to the size of the area considered and the amount of data available. 

Thanks to the increasing use of mobile phones as data sources, larger scale studies with a focus on air transportation have been possible. In the United States, Marzuoli et al. \cite{marzuoli2019ImplementingValidatingAir} implement and validate a method to detect domestic air passengers using mobile phone data available on a nationwide scale. Though the main focus of this work is the passenger behavior at airports, the granularity of the data facilitates analysis of each phase within the full door-to-door trip. Marzuoli et al. \cite{marzuoli2018PassengercentricMetricsAir} then combine mobile phone data with social media data to analyze passenger experiences in airports during a major disruptions. In Europe, within the BigData4ATM project\footnote{\url{www.bigdata4atm.eu}}, Garcia-Albertos et al. \cite{garcia-albertos2017UnderstandingDoortoDoorTravel} also present a methodology for measuring door-to-door travel times using mobile phone data, illustrated through trips between Madrid and Barcelona. Mobile phone data are, however, proprietary data, which are difficult to access for research.

Grimme and Martens \cite{grimme2019Flightpath2050Revisited} propose a model analyzing the feasibility of the 4-hour goal proposed by FlightPath 2050 based on airport-to-airport flight times and a simplified model of access to and from airports. Sun et al. \cite{sun2018CompetitivenessOndemandAir} use open source maps and datasets to calculate door-to-door minimum travel time estimations in order to study the possible competitiveness of air taxis. 

In the upcoming sections, the model and analysis presented are also based on already available online data but with a post operation approach. The aim of this model is to create a method to measure the average door-to-door travel times once trips are completed to analyze and compare  available modes of transportation. We have applied the first version of this method to two intra-European multi-modal trips, thus comparing air transportation and rail transportation \cite{monmousseau2019DoortodoorTravelTime}. We then used an improved version of the same method, leveraging four different data sources (ride-sharing, flight, phone, and census data) and adapted to the conditions in the United States, to compare trips using direct flights between five US cities, three of them on the West Coast and the other two on the East Coast \cite{monmousseau2020DoortodoorAirTravel}. 

In this paper, we offer a data-driven model for the computation of door-to-door travel times that harnesses recently available data along with public data. The data-driven methods developed can be applied for most multi-modal trips between two cities where relevant data are available and are not limited to the air transportation system. The range of new analyses available using this model is illustrated with multiple modal analysis of an intra-European trip, a per-leg analysis of multiple intra-USA trips and an analysis of the impact of severe weather disruptions. These analyses have direct applications for passengers, urban planners and decision makers and highlight the difference between taking a flight-centric approach to the air transportation system and taking a passenger-centric approach.

Section\,\ref{sec:model} of this paper presents the data-driven, full door-to-door travel time model; Section\,\ref{sec:appli_paris} showcases a first set of analyses and applications facilitated by this model for trips between Amsterdam and Paris; and, Section\,\ref{sec:appli_us} focuses on a set of analyses for trips within the United States where more data are available; Finally, Section\,\ref{sec:concl} concludes this paper and proposes future research directions.

\section{The full door-to-door data-driven model}
\label{sec:model}
\begin{figure}[ht]
\centering
\begin{tikzpicture}
\node[] (q0) {};
\node [draw,fill=gray!20,align=center, rounded corners, right of=q0,opacity=1] (dep) {Departure\\ station};
\node[right of=dep] (qi) {};
\node [draw,fill=gray!20,align=center, rounded corners, right of=qi,opacity=1] (arr) {Arrival\\ station};
\node[right of=arr] (qf) {};
\draw[opacity=1] (q0) edge node {$t_\text{to}$} (dep);
\draw[opacity=1] (dep) edge[loop above] node {$t_\text{dep}$} (dep);
\draw[opacity=1] (dep) edge[bend left=45] node {$t_\text{in}$} (arr);
\draw[opacity=1] (arr) edge[loop above] node {$t_\text{arr}$} (arr);
\draw[opacity=1] (arr) edge node {$t_\text{from}$}  (qf);
\end{tikzpicture}
\caption{Model of the full door-to-door travel time.}
\label{fig:door2door_model}
\end{figure}
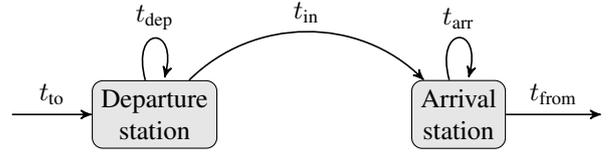

Similarly to \cite{garcia-albertos2017UnderstandingDoortoDoorTravel} and \cite{sun2018CompetitivenessOndemandAir}, we can deconstruct the travel time $T$ for trips with direct flights or direct train links into five different trip phases, represented in Figure\,\ref{fig:door2door_model} and summarized in equation \eqref{eq:door2door},
\begin{equation}
\label{eq:door2door}
T = t_\text{to} + t_\text{dep} + t_\text{in} + t_\text{arr} + t_\text{from} ~,
\end{equation}
where
\begin{itemize}
\item $t_\text{to}$ is the time spent traveling from the start of the journey to the departure station (e.g. train station or airport),
\item $t_\text{dep}$ is the time spent waiting and going through security processes (if any) at the departure station,
\item $t_\text{in}$ is the time actually spent in flight or on rails,
\item $t_\text{arr}$ is the time spent at the arrival station (e.g. going through security processes),
\item $t_\text{from}$ is the time spent traveling from the arrival station to the final destination.
\end{itemize}

The full model for door-to-door travel time proposed in this paper is established by data-driven methods used to calculate the values of the different times contained in equation \eqref{eq:door2door}. These data-driven methods are described in Sections \ref{sec:model_uber} through \ref{sec:model_total}.

This study focuses on air and rail transportation as main transportation modes, which give the value of $t_\text{in}$, though the process can also be applied to inter-city bus trips. Furthermore, it is assumed that passengers travel by road when arriving or leaving the main station (airport or train station) for the calculation of $t_\text{to}$ and $t_\text{from}$.
In response to data availability, the case studies only consider six major US cities (Atlanta, Boston, Los Angeles, Seattle, San Francisco and Washington D.C.) and two European capitals (Amsterdam and Paris).

\subsection{Travel time from the origin location to the departure station and from the arrival station to the final destination}
\label{sec:model_uber}
We can estimate the road transit times from origin location to departure station ($t_\text{to}$) and from arrival station to final destination ($t_\text{from}$) by using aggregated and publicly available data from taxi or ride-sharing services.
Uber \cite{UberWebsite} is a ride-sharing service launched in 2010 and located in major urban areas on six continents; it has recently released anonymized and aggregated travel time data for certain of the urban areas where it operates. The available data consist of the average, minimum and maximum travel times between different zones (e.g. census tracts in the case of US cities) within serviced area from all Uber rides aggregated over five different periods for each considered day. The five considered periods, used throughout this study, are defined as follows:
\begin{itemize}
\item Early Morning: from midnight to 7am
\item AM: from 7am to 10am
\item Midday: from 10am to 4pm
\item PM: from 4pm to 7pm
\item Late Evening: from 7pm to midnight
\end{itemize}
There are days when the travel times between some zones are only aggregated at a daily level. Travel times are associated with their mean starting door time, i.e. the mean of all the time stamps from the trip contained in the zone of departure.

Since Uber was initially introduced in the US, the impact of Uber in US urban transit has already been the focus of several studies prior to this data release.
Li et al. \cite{li2016OndemandRidesharingServices} concludes that, at an aggregated level, Uber tends to decrease congestion in the US urban areas where it was introduced. Later, Erhardt et al. \cite{erhardt2019TransportationNetworkCompanies} build a model showing that ride sharing companies do increase congestion using the example of San Francisco. Hall et al. \cite{hall2018UberSubstituteComplement} focus on whether Uber complemented or substituted public transit by studying the use of public transit system before and after Uber's entry date in different US cities. Wang and Mu \cite{wang2018SpatialDisparitiesUber} study Uber's accessibility in Atlanta, GA (US) by using the average wait time for a ride as a proxy and conclude that the Uber use is not associated to a specific social category. Following the release of Uber data, Pearson et al. \cite{pearson2018TrafficFlowAnalysis} propose a traffic flow model based on this aggregated Uber data and use it to analyze traffic patterns for seven cities world-wide. Assuming Uber rides as part of the road traffic flow, this study considers that Uber's travel times are an acceptable proxy of the actual travel times by road. In cities where busses don't have specific road lanes, these travel times are a valid proxy for both car and bus trips. This paper limits its scope to road access to and egress from considered stations. The analysis of subway alternatives is not considered in this paper.

Each US city is divided into their census tracts; Paris into the IRIS zones used by INSEE \cite{InseeWebsite} for census, and Amsterdam into its official districts called \emph{wijk}. 

\subsection{Dwell time at stations}
\label{sec:model_wait}
The dwell time at a station, either $t_\text{dep}$ or $t_\text{arr}$, is defined as the time spent at the station, whether going through security processes, walking through the station, or waiting. The time spent at each station depends on the mode considered, the specific trip, and whether the passenger is departing or arriving. The dwell time at departure can be split into two components, 
\begin{equation}
\label{eq:split_dep}
t_\text{dep} = t_\text{sec} + t_\text{wait}. ~,
\end{equation}
a processing time, $t_\text{sec}$, necessary to get through security (if any) and through the station to the desired gate or track, and an extra wait time, $t_\text{wait}$, due to unanticipated delays.

Processing times at US airports are based on the average wait times at airports extracted from the study of Marzuoli et al. \cite{marzuoli2019ImplementingValidatingAir}. The six US airports under study in this paper are: Hartsfield-Jackson Atlanta International Airport (ATL), Boston's Logan International Airport (BOS) and Ronald Reagan Washington National Airport (DCA) for the East Coast, Los Angeles International Airport (LAX), Seattle-Tacoma International Airport (SEA) and San Francisco International Airport (SFO) for the West Coast. Processing times at European airports are assumed invariant between airports and determined using most airline recommendations. The three European airports under study are: Paris Charles de Gaulle Airport (CDG), Paris Orly Airport (ORY) and Amsterdam Airport Schiphol (AMS).

The average dwell times at these airports are summarized in Table\,\ref{tab:t_at_mode_us} for US airports and in Table\,\ref{tab:t_at_mode_eu} for European airports.
\begin{table}[h!]
\centering
\caption{Average dwell time spent at US airports in minutes.}
\label{tab:t_at_mode_us}
\begin{tabular}{|l|c|c|c|c|c|c|}
\hline
   &  \textbf{ATL} & \textbf{BOS} & \textbf{DCA} & \textbf{LAX} & \textbf{SEA} & \textbf{SFO}  \\ \hline
Time at departure & 110 & 105 & 100 & 125 & 105 & 105  \\
Time at arrival & 60 & 40   & 35   & 65   & 50   & 45 \\
\hline
\end{tabular}
\vspace{1em}
\caption{Average dwell time spent at European airports in minutes.}
\label{tab:t_at_mode_eu}
\begin{tabular}{|l|c|c|c|}
\hline
   & \textbf{AMS} & \textbf{CDG} &  \textbf{ORY} \\ \hline
Time at departure &  90 & 90 & 90  \\
Time at arrival & 45 & 45 & 45 \\
\hline
\end{tabular}
\end{table}

With regard to processing times at train stations, based on the recommendation of the train station websites, the departure dwell time is set at 15 minutes and the arrival dwell time is set at 10 minutes for all train stations. We can improve these estimates by gathering data from GPS or mobile phone sources as well as WiFi beacons within airports and train stations, and by using a method similar to Nikoue et al. \cite{nikoue2015PassengerFlowPredictions}.

We can calculate the extra wait times when the scheduled and actual departure or arrival times are available. For US airports, these wait times are calculated only for departure using the publicly available data from the Bureau of Transportation Statistics (BTS) \cite{BTSwebsite}. They were obtained by subtracting the scheduled departure time from the actual flight departure time.

\subsection{Time in flight or on rail}
\subsubsection{US flights}
\label{sec:model_us_in}
The actual flight time was calculated based on the data from BTS using the actual departure/arrival times of all direct flights between each city pairs from January 1\st\, 2018 to March 31\st\, 2018. Cancelled flights are not considered in this study and were discarded.

\subsubsection{European trips}
\label{sec:model_eu_in}
In Europe, we assume that flights and trains are on time and follow a weekly schedule, due to a lack of publicly centralized flight schedule data. The weekly schedules are extracted from actual train and flight schedules gathered over a period of several months and are assumed applicable over the full period under study.

\subsection{Full door-to-door travel time}
\label{sec:model_total}
Our model assumes that travelers plan their departure time to arrive at the departure station exactly $t_\text{sec}$ minutes (eq. \eqref{eq:split_dep}) before the scheduled departure time of their flight or train. We use this assumption to determine the value of $t_\text{to}$ since it defines the period of the day to consider when extracting the Uber average time from the origin location to the departure station. We extract the value of $t_\text{from}$ by using the actual arrival time of the flight or train. When only daily aggregated times are available in the Uber data, these times are used for each period of the day in proxy.  

\section{Flights versus trains: a comparison of different access modes to Paris}
\label{sec:appli_paris}
Let us consider a traveler leaving from Amsterdam city center to reach the Paris area. We have chosen the city center of Amsterdam as it covers both tourists and business travelers, but the proposed door-to-door travel time model and subsequent analysis can be applied from any zone. Three possible means of transportation are under study for this trip: plane from Amsterdam Airport Schiphol to Paris Charles De Gaulle airport (CDG) or via Paris Orly (ORY), or train from Amsterdam Centraal via Paris Gare du Nord (GDN).

\definecolor{green1}{RGB}{0,147,156}
\definecolor{green2}{RGB}{93,186,191}
\definecolor{green3}{RGB}{186,225,226}
\definecolor{red1}{RGB}{248,192,170}
\definecolor{red2}{RGB}{221,119,85}
\definecolor{red3}{RGB}{194,46,0}

\subsection{Flight and train schedules}
As in Section\,\ref{sec:model_eu_in}, the flight and train schedules used for this study are weekly schedules. The weekly flight schedules between Amsterdam Airport Schiphol (AMS) and CDG or ORY were extracted from the actual flight schedules from December 2019 to January 2020, and it was assumed that the obtained weekly schedules would run from January 1\st\, 2018 to September 30\nth\, 2019. These weekly schedules are summarized in Table\,\ref{tab:sched_ams2ory} for flights between AMS and ORY, and in Table\,\ref{tab:sched_ams2cdg} for flights between AMS and CDG.

\begin{table}[h!t]
\parbox[t]{.49\textwidth}{
\centering
\caption{Extracted weekly schedule from Amsterdam to Paris via ORY.}
\label{tab:sched_ams2ory}
\begin{filecontents*}{tables/ams2ory.csv}
Mon,Tue,Wed,Thu,Fri,Sat,Sun,crs_dep_time,crs_arr_time
x,x,x,x,,,,10:25,11:45
,,,,x,,,14:45,16:05
x,x,x,x,x,,,18:50,20:10
,,,,,,x,19:40,21:00
\end{filecontents*}
\csvautotabular[table head=\hline\bfseries Mo & \bfseries Tu & \bfseries We & \bfseries Th & \bfseries Fr & \bfseries Sa & \bfseries Su & \bfseries Ams. & \bfseries Paris\\\hline]{tables/ams2ory.csv}
}\hfill
\parbox[t]{.49\textwidth}{
\centering
\caption{Extracted weekly schedule from Amsterdam to Paris via CDG.}
\label{tab:sched_ams2cdg}
\begin{filecontents*}{tables/ams2cdg.csv}
Mon,Tue,Wed,Thu,Fri,Sat,Sun,crs_dep_time,crs_arr_time
x,x,x,x,x,x,x,06:50,08:10
x,x,x,x,x,x,x,07:20,08:45
x,x,x,x,x,x,x,08:10,09:35
x,x,x,x,x,x,x,09:30,10:55
x,x,x,x,x,x,x,10:20,11:45
x,x,x,x,x,x,x,12:25,13:40
x,x,x,x,x,x,x,13:55,15:15
x,x,x,x,x,x,x,14:50,16:10
x,x,x,x,x,x,x,16:35,17:50
x,x,x,x,x,x,x,17:45,19:00
x,x,x,x,x,,x,19:10,20:30
x,x,x,x,x,x,x,20:25,21:45
\end{filecontents*}
\csvautotabular[table head=\hline\bfseries Mo & \bfseries Tu & \bfseries We & \bfseries Th & \bfseries Fr & \bfseries Sa & \bfseries Su & \bfseries Ams. & \bfseries Paris\\\hline]{tables/ams2cdg.csv}
}
\end{table}

The weekly train schedule between Amsterdam Centraal station and Paris Gare du Nord (GDN) is similarly extracted from the actual train schedule of the year 2019 and applied to the year 2018. It is summarized in Table\,\ref{tab:sched_ams2gdn}. Night trains are not considered for this study.

\begin{table}[h!t]
\centering
\caption{Extracted weekly schedule from Amsterdam to Paris via GDN.}
\label{tab:sched_ams2gdn}
\begin{filecontents*}{tables/amc2gdn.csv}
Mon,Tue,Wed,Thu,Fri,Sat,Sun,crs_dep_time,crs_arr_time
x,x,x,x,x,,x,06:15,09:35
x,x,x,x,x,x,x,07:15,10:38
x,x,x,x,x,x,x,08:15,11:35
x,x,x,x,x,x,,09:15,12:35
,,,,,x,x,10:15,13:38
x,x,x,x,x,x,x,11:15,14:35
x,x,x,x,x,x,x,13:15,16:38
x,x,x,x,x,,x,14:15,17:35
x,x,x,x,x,x,x,15:15,18:35
x,x,x,x,x,,x,16:15,19:35
x,x,x,x,x,,x,17:15,20:35
x,x,x,x,x,x,x,18:15,21:38
x,x,x,x,x,,x,19:15,22:35
,,,,,,x,20:15,23:38
\end{filecontents*}
\csvautotabular[table head=\hline\bfseries Mo & \bfseries Tu & \bfseries We & \bfseries Th & \bfseries Fr & \bfseries Sa & \bfseries Su & \bfseries Ams. & \bfseries Paris\\\hline]{tables/amc2gdn.csv}
\end{table}

These schedules already highlight the major differences between the three considered modes. The flight schedule through ORY contains the fewest possibilities, limited to two flights daily, whereas the other two models offer hourly scheduled transport. Another notable difference can be seen with respect to the station-to-station travel times: flights between Amsterdam and Paris (both CDG and ORY) take 1h20 ($\pm$ 5 minutes) while train rides between Amsterdam Centraal and Paris GDN take 3h20 ($\pm$ 3 minutes).

\subsection{Average total travel time mode comparison}
\label{sec:appli_bestmode}
The proposed data-driven door-to-door model can be used to evaluate and compare the range of each considered mode, which helps to understand better the urban structure and behavior from a transportation point of view. 

For each trip $\tau$ (flight or rail) over the period $\mathcal{D}$ from January 1\st\,2018 to September 30\nth\,2019, the associated average full door-to-door travel time $\bar{T}(\tau)$,
\begin{equation}
\label{eq:avgd2d_eu}
\bar{T}(\tau) = \bar{t}_\text{to}(\tau) + t_\text{dep}(\tau) + t_\text{in}(\tau) + t_\text{arr}(\tau) + \bar{t}_\text{from}(\tau)~,
\end{equation}
where $\bar{t}_\text{to}(\tau)$ is the average ride time between Amsterdam city center and the departure station (AMS or Amsterdam Centraal) for the trip $\tau$, $\bar{t}_\text{from}(\tau)$ is the average ride time from the arriving station (CDG, ORY or GDN) to the final arrival zone for the trip $\tau$.

The same daily periods as those used in the Uber data (see Section\,\ref{sec:model_uber}) are considered here to categorize the trips into five groups depending on the time of arrival at the final destination. For each day $d$ and each period $p$, the mean per arrival zone $z$ of the average door-to-door travel times is calculated for each mode $m$,
\begin{equation}
\label{eq:mavgd2d_eu}
E^{d,p}_z(m) = \frac{1}{|\mathcal{T}^{d,p}_z|}\sum_{\tau \in \mathcal{T}^{d,p}_z} \bar{T}(\tau)~,
\end{equation}
where $\mathcal{T}^{d,p}_z$ is the set of flight and rail trips that end at zone $z$ on day $d$ and period $p$.

For each day $d$, each period $p$ and each arrival zone $z$, the mode with the shortest mean travel time $E^{d,p}_z(m)$ is kept. The number of times $N^p_z(m)$ a mode $m$ has had the shortest mean travel time is counted for each zone $z$ and for each period $p$ over the twenty-one month period $\mathcal{D}$,
\begin{equation}
\label{eq:numbestd2d_eu}
N^p_z(m) = | \{ d\in \mathcal{D} ~|~ m = \text{arg}\min_{m_1} E^{d,p}_z(m_1) \} | ~.
\end{equation}

This distribution of modes over the different zones can help travelers choose the mode of transport that is best suited depending on the desired arrival zone and on the desired time of arrival. It can also help urban planners to better understand the road network linking the different stations to the city.

\begin{figure}[h!t]
\centering
\begin{subfigure}[b]{.49\textwidth}
\includegraphics[width=\textwidth]{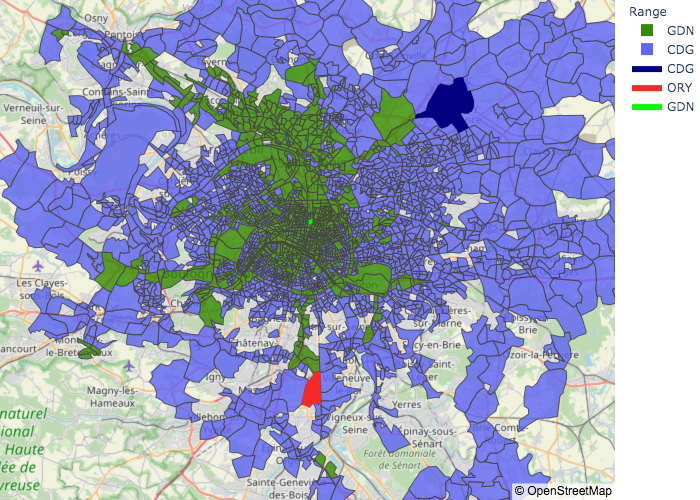}
\caption{Morning (AM)}
\end{subfigure}
\begin{subfigure}[b]{.49\textwidth}
\includegraphics[width=\textwidth]{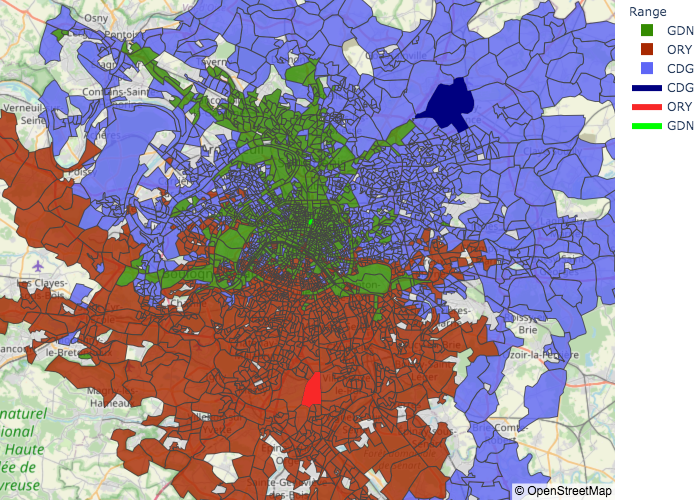}
\caption{Midday}
\end{subfigure}
\caption{Comparison of the average total travel times to the Paris area between the three considered arrival stations (CDG: blue, ORY: red, GDN: green) for a trip starting from Amsterdam city center for different trip termination periods.}
\label{fig:paris_best_cmp}
\end{figure}

Figure\,\ref{fig:paris_best_cmp} shows the fastest mode to reach the different zones in the Paris dataset for the five different periods of the days used by the Uber dataset. For each zone $z$ and each period $p$, the fastest mode associated is the mode $m$ having the highest $N^p_z(m)$, i.e. the highest number of days with the lowest average total travel time over the considered date range. The zones best reached through CDG are indicated in blue, ORY in red and GDN in green. 

We can draw several conclusions from these maps. The absence of zones best reached by plane via ORY (in red) is particularly noticeable in the morning period: An important area of South-West Paris is not reached by Uber rides neither from GDN nor from CDG. These maps would advocate for an increase in frequency for the AMS-ORY flights from a traveler's perspective.

From a structural perspective, the interstate linking Paris to CDG is visible on all three maps since it enables travelers through GDN to reach zones close to CDG faster than if they flew to CDG directly. The perimeter highway circling Paris is also a major aid to GDN and is visible on the maps where there is an important competition between GDN and CDG. The section of the perimeter highway farthest from GDN (i.e. in the south-west of Paris) is, however, overtaken by either airport depending on the period of the day. The rest of GDN influence zone is fairly invariant from a period to another.

Using a similar map representation for trips ending in the afternoon but not shown here for space considerations,  ORY's range is limited during the afternoon, with CDG taking over some zones close to ORY. This is essentially due to the limited number of flights landing in the afternoon (one per week, every Friday) compared to the daily arrival of CDG flights.

\subsection{Average total travel time distribution analysis}
\label{sec:appli_time}
Once the fastest mode to reach each zone is determined, it is possible to analyze the fastest full door-to-door travel time for each zone. This approach gives an overview of the level of integration of airports, train stations, and road structure and can indicate zones that are less reachable than others and would thus require more attention from urban planners.

The fastest full door-to-door travel time associated with a zone $z$ at a period $p$ is calculated as the average fastest travel time to reach zone $z$ at period $p$ across all modes and over the period $\mathcal{D}$,
\begin{equation}
\label{eq:avgfasttime_eu}
\bar{E}_z^p =  \frac{1}{|\mathcal{D}|} \sum \min_m E_z^{d,p}(m) ~.
\end{equation}

\begin{figure*}[h!t]
\centering
\begin{subfigure}[b]{.49\textwidth}
\includegraphics[width=\textwidth]{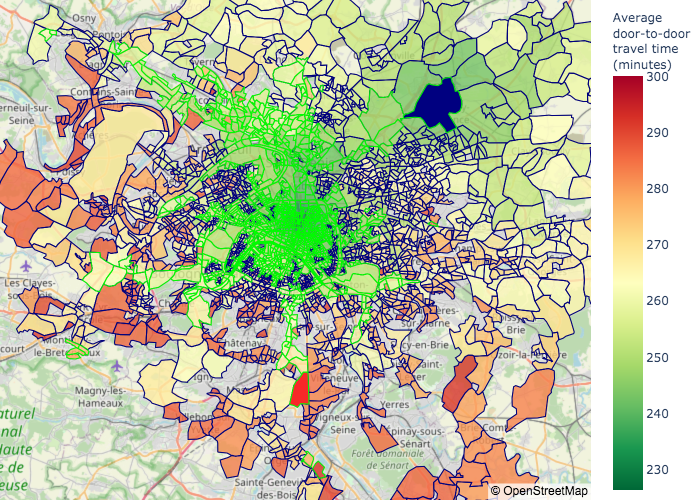}
\caption{Morning (AM)}
\end{subfigure}
\begin{subfigure}[b]{.49\textwidth}
\includegraphics[width=\textwidth]{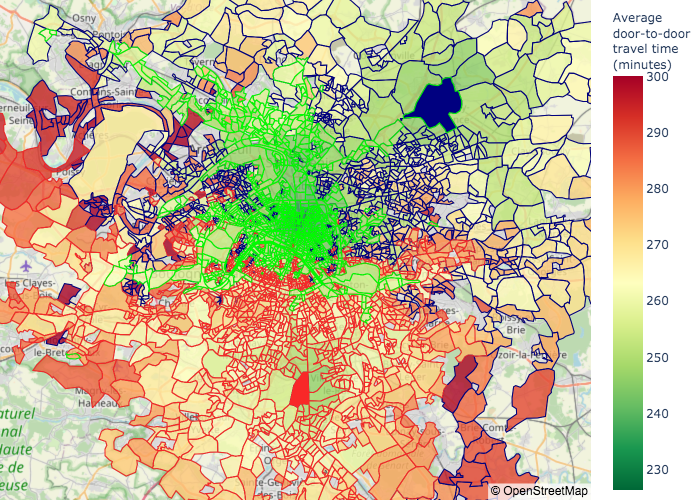}
\caption{Midday}
\end{subfigure}
\caption{Comparison of the fastest full door-to-door travel times to the Paris area between the three considered arrival stations (CDG: blue, ORY: red, GDN: green) for a trip starting from Amsterdam city center for different trip termination periods. The contour color of each zone indicates the fastest mode to reach it.}
\label{fig:paris_time_cmp}
\end{figure*}

Figure\,\ref{fig:paris_time_cmp} displays the fastest full door-to-door travel times $\{\bar{E}_z^p\}_{p,z}$ to reach the different zones in the Paris dataset for trips finishing in the morning or at midday. The color scale that indicates the fastest full door-to-door travel times is identical in all subfigures. The contour of each zone indicates the fastest mode to reach it according to the results from Section\,\ref{sec:appli_bestmode} using the same color code as Figure\,\ref{fig:paris_best_cmp}, i.e. the zones reached faster through CDG are surrounded in blue, ORY in red and GDN in green.

For a better comparison, the distribution of the number of zones per period reached within four time intervals (less than 4 hours, between 4 hours and 4 hours and 30 minutes, between 4 hours and 30 minutes and 5 hours, and more than 5 hours) is presented in Table\,\ref{tab:paris_count}.
\begin{table}[htb]
\centering
\caption{Number of zones per mode and period of the day grouped by full door-to-door travel time intervals. The original dataset is the same as that used to generate Figure\,\ref{fig:paris_time_cmp}.}
\label{tab:paris_count}
\begin{tabular}{|c|c|c|c|c|c|c|c|}
\hline
\bfseries Mode & \bfseries Time interval & \bfseries Early & \bfseries AM & \bfseries Midday & \bfseries PM & \bfseries Late \\\hline 
\multirow{4}{*}{CDG}&$t\leq$ 4h&0&4&6&5&11\\
&4h $<t\leq$ 4h30&22&1306&866&1189&845\\
&4h30 $<t\leq$ 5h&0&653&433&498&384\\
&$t>$ 5h&187&0&15&13&11\\\hline
\multirow{4}{*}{GDN}&$t\leq$ 4h&398&247&247&290&314\\
&4h $<t\leq$ 4h30&775&818&731&719&641\\
&4h30 $<t\leq$ 5h&0&14&8&6&8\\
&$t>$ 5h&0&0&0&0&0\\\hline
\multirow{4}{*}{ORY}&$t\leq$ 4h&0&0&0&0&0\\
&4h $<t\leq$ 4h30&0&0&906&563&997\\
&4h30 $<t\leq$ 5h&0&0&397&259&425\\
&$t>$ 5h&0&0&0&0&1\\

\hline
\end{tabular}
\end{table}

We see a dissymmetry between the north and the south of Paris by looking only at the time color scale in Figure\,\ref{fig:paris_time_cmp}. The number of green zones, i.e. zones reachable in less than 4 hours and 20 minutes, and the surface covered by these green zones are more important in the northern part of Paris than in the southern part of Paris, including in areas close to Paris Orly airport. Combining this observation with the contour color of each zone suggests that Paris Orly airport is not as well integrated in the Parisian road structure as Paris Charles de Gaulle airport, which can make it less attractive for travelers desiring to travel by air and thus less competitive.

We can complete a more quantitative analysis from Table\,\ref{tab:paris_count}, with some of the main findings listed here: 
\begin{itemize}
\item Only 10\% of the arrival zones are reached in less than 4 hours from Amsterdam city center.
\item Zones that can be reached in less than 4 hours from Amsterdam city center are overwhelmingly reached by train through Paris Gare du Nord (98\%). 
\item A trip from Amsterdam city center to Paris going through ORY always takes more than 4 hours.
\item 78\% of the arrival zones are reached in less than 4 hours and 30 minutes from Amsterdam city center when combining all three possible modes.
\end{itemize}
Therefore, we can use these results to assess how well the 4-hour goal from ACARE FlightPath 2050 is engaged.

\subsection{Reliability issues}
The proposed full door-to-door travel time model assumes that passengers choose their departure time in order to arrive exactly $t_\text{sec}$ before the scheduled departure of their plane or train and that they also know how long it takes to reach the departure station. However, in reality, there is an uncertainty in the time the traveler will spend reaching the airport and in airport processing times. This uncertainty often leads to an additional buffer time implying an earlier departure time for the traveler. Using the presented model with the available data, we can find the most reliable mode to use per arrival zone. The most reliable mode for a given arrival zone is defined as the mode with the lowest variability in travel time, i.e. the mode where the difference between the maximum travel time and the minimum travel time to reach that zone is the lowest. This comparison is useful for passengers or trips that require an accurate arrival time rather than a minimum travel time.

\begin{figure}[h!t]
\centering
\begin{subfigure}[b]{.49\textwidth}
\includegraphics[width=\textwidth]{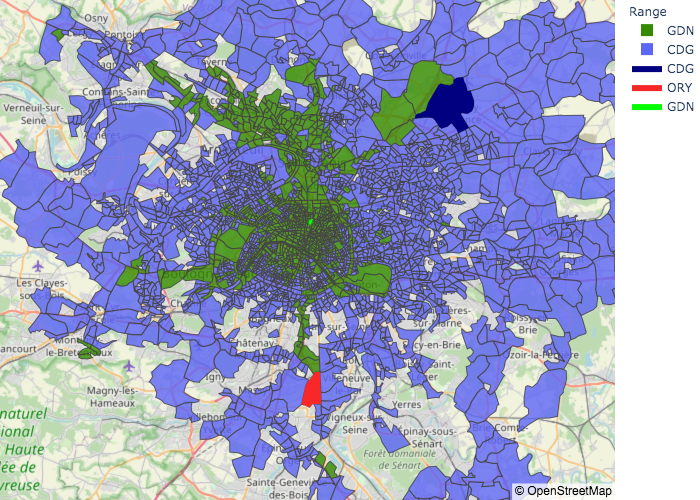}
\caption{Morning (AM)}
\end{subfigure}
\begin{subfigure}[b]{.49\textwidth}
\includegraphics[width=\textwidth]{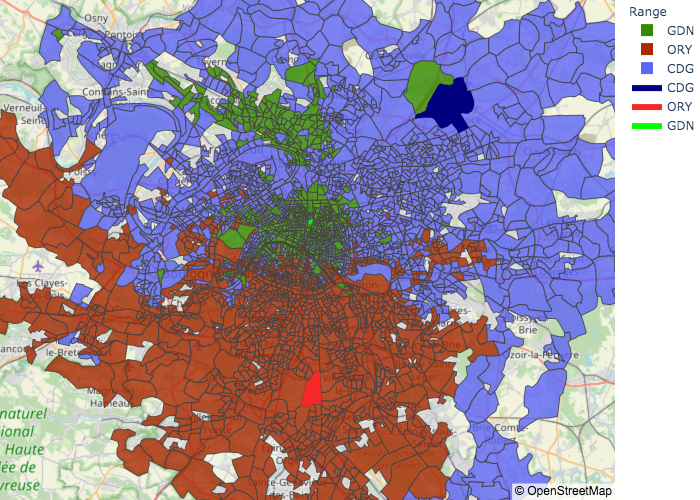}
\caption{Midday}
\end{subfigure}
\caption{Comparison of the average variability of travel times to the Paris area between the three considered arrival stations (CDG: blue, ORY: red, GDN: green) for a trip starting from Amsterdam city center for different trip termination periods.}
\label{fig:paris_safest_cmp}
\end{figure}

Figure\,\ref{fig:paris_safest_cmp} shows the most reliable mode on average to reach the different zones in the Paris dataset for trips finishing in the morning or at midday. As for the previous analysis, the period was determined using the departure time of the full door-to-door trip and uses the same color code, i.e. the zones reached most reliably through CDG are indicated in blue, ORY in red and GDN in green. For each zone and each period of the day, the most reliable mode associated is the mode having the highest number of days with the lowest average variability travel time over the considered date range. 

Though Figure\,\ref{fig:paris_safest_cmp} and Figure\,\ref{fig:paris_best_cmp} are similar, there are some major differences between average efficiency and average reliability. For example, though it is on average faster to reach the zones close to the highway leading to CDG by train, after 10:00 it is safer from a time variability perspective to reach them via CDG. From a reliability perspective, CDG has claimed the quasi totality of the zones surrounding it, except in the early morning where trips through GDN are still better. When we compare all three modes using this metric, it appears that GDN has the greatest decrease in competitiveness, with its range smaller than when considering average travel times.

\subsection{Impact of faster processing times}
The major difference between air and rail travel is the necessary processing time both at departure and at arrival. In this particular study, with a flight time of about 80 minutes, the current assumption of a departure processing time of 90 minutes implies that travelers spend more time at their departure airport than in flight, which greatly impacts the rapidity of air travel. With the presented model, one can modify these assumed processing times in order to study the impact of improving these times both from an airport perspective and a passenger perspective. Let's assume that the processing time at airports is improved from 90 to 60 minutes at departure and from 45 to 30 minutes at arrival. These modifications could be achieved in reality considering that this is an intra-Schengen trip and that there isn't any border control.

\begin{figure}[h!t]
\centering
\begin{subfigure}[b]{.49\textwidth}
\includegraphics[width=\textwidth]{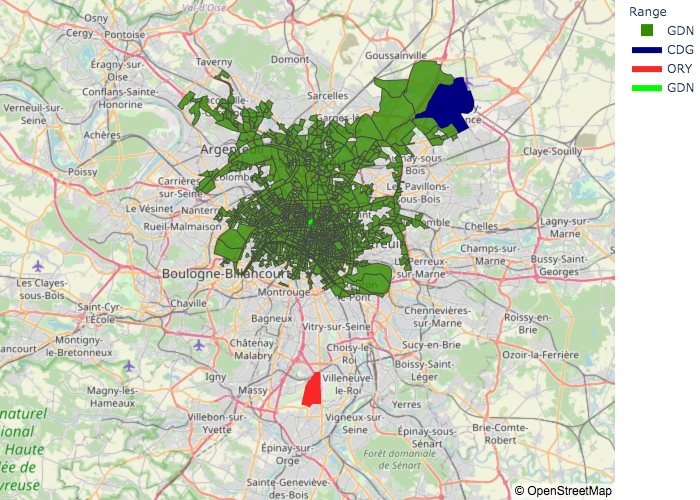}
\caption{Early morning with faster processing times}
\label{fig:paris60_best_early}
\end{subfigure}
\begin{subfigure}[b]{.49\textwidth}
\includegraphics[width=\textwidth]{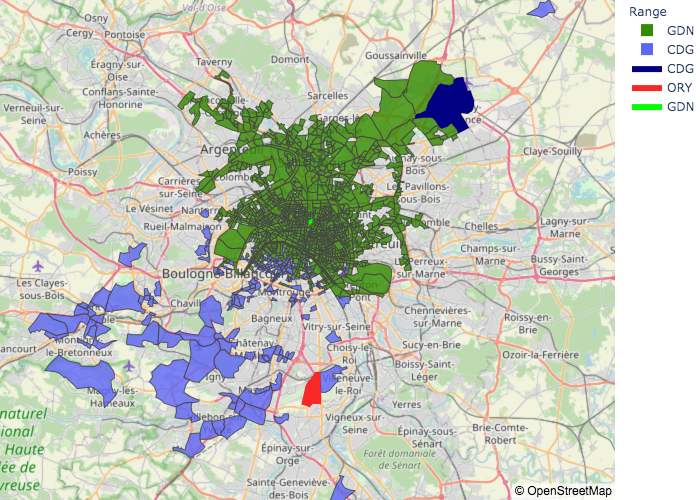}
\caption{Early morning with normal processing times}
\label{fig:paris90_best_early}
\end{subfigure}
\caption{Comparison of the average total travel times to the Paris area assuming faster airport processing times between the three considered arrival stations (CDG: blue, ORY: red, GDN: green) for a trip starting from Amsterdam city center for trips arriving at destination early in the morning. The corresponding map with normal processing times is also reproduced.}
\label{fig:paris60_time_cmp_early}
\end{figure}
\begin{figure*}[h!t]
\centering
\begin{subfigure}[b]{.49\textwidth}
\includegraphics[width=\textwidth]{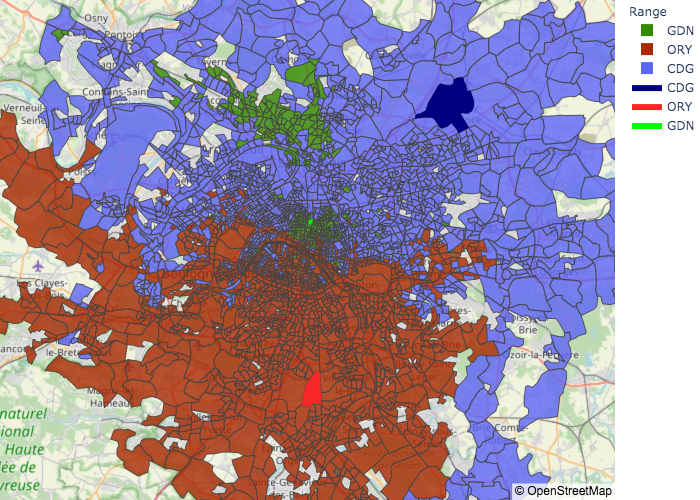}
\caption{Midday with faster processing times}
\label{fig:paris60_best_midday}
\end{subfigure}
\begin{subfigure}[b]{.49\textwidth}
\includegraphics[width=\textwidth]{paris_best_midday.png}
\caption{Midday with normal processing times}
\label{fig:paris90_best_midday}
\end{subfigure}
\caption{Comparison of the average total travel times to the Paris area assuming faster airport processing times between the three considered arrival stations (CDG: blue, ORY: red, GDN: green) for a trip starting from Amsterdam city center for trips arriving at destination at midday. The corresponding maps with normal processing times are also reproduced.}
\label{fig:paris60_time_cmp_midday}
\end{figure*}

Figures\,\ref{fig:paris60_time_cmp_early}\&\,\ref{fig:paris60_time_cmp_midday} show which is the fastest mode on average to reach the different zones in the Paris dataset for trips arriving at destination early in the morning or at midday. 

The first major difference with this processing time improvement can be seen for trips arriving in the early morning (Figure\,\ref{fig:paris60_time_cmp_early}): all zones previously reached through CDG are no longer accessed at this period since they were associated to the 21:45 flight of the previous day. This indicates that all trips through CDG start and end on the same day, with no trips finishing after midnight.

Looking at trips arriving at midday (Figure\,\ref{fig:paris60_time_cmp_midday}), trips through CDG are greatly advantaged by this time improvement, with CDG taking over more than half of GDN's previous influence zone. This range increase from CDG can be explained both by faster door-to-door travel times and by the increase of trips through CDG arriving at midday (rather than in the afternoon). As a matter of fact, besides in the early morning, GDN loses its competitiveness against both airports, with its range greatly shrinking in size. The competition between CDG and ORY remains unchanged, which is understandable since they both received the same processing time improvement.

A quantitative analysis similar to the analysis presented in Table\,\ref{tab:paris_count} concludes that all trips are now conducted in less than five hours and that 99.8\% of the zones reachable are reached in less than 4h30. ORY sees some major improvements with 97.5\% of the zones best reached through it reached in less than four hours (compared to no trips in less than 4h in the initial model), while increasing the number of zones it reaches the fastest.

Using a map representation similar to Section\,\ref{sec:appli_time}, but not presented here due to space considerations, it is possible to notice a 20-30 minutes shift in the time distribution for every period except for early morning trips since train processing times were unchanged. The upper bound travel time is also unchanged for trips arriving in the morning, which would indicate that for some zones, the processing time improvement resulted in no improvement or even a worsening of the full trip travel time. Besides that exception, in this case a 45 minutes improvement in airport processing time leads only to a maximum of 30 minutes of average total travel time improvement due to the influence of train trips through GDN.

\section{A multi-modal analysis of the US air transportation system}
\label{sec:appli_us}
Additional insights are gained from this full door-to-door travel time model thanks to the availability of complementary data. The United States is a federal state the size of a continent, therefore various aggregated and centralized datasets are more easily available to all. Several of these datasets are used in this section to add applications to the presented full door-to-door model. This US study is limited to the period from January 1st 2018 to March 31st 2018.

\subsection{Flight schedule}
As presented in the model definition in Section\,\ref{sec:model_us_in}, both the scheduled flight times and the actual flight schedules of most domestic flights can be obtained via the Department of Transportation Bureau of Transportation Statistics (BTS) \cite{BTSwebsite}. This study considers only the six US airports presented in Section\,\ref{sec:model_wait}, three East-coast airports - Hartsfield-Jackson Atlanta International Airport (ATL), Boston's Logan International Airport (BOS) and Ronald Reagan Washington National Airport (DCA) - and three West-coast airports - Los Angeles International Airport (LAX), Seattle-Tacoma International Airport (SEA) and San Francisco International Airport (SFO).

During this three-month period, 38,826 flights were considered, which corresponds to 3,523 early flights, 8,170 morning flights, 13,451 midday flights, 6.695 afternoon flights and 6,987 evening flights.
The full door-to-door travel times were then calculated for each scheduled flight from January 1\st, 2018 to March 31\st, 2018, using the model presented in Section\,\ref{sec:model}.

\subsection{Leg analysis}
\begin{figure}[h!bt]
\centering
\includegraphics[width=\columnwidth]{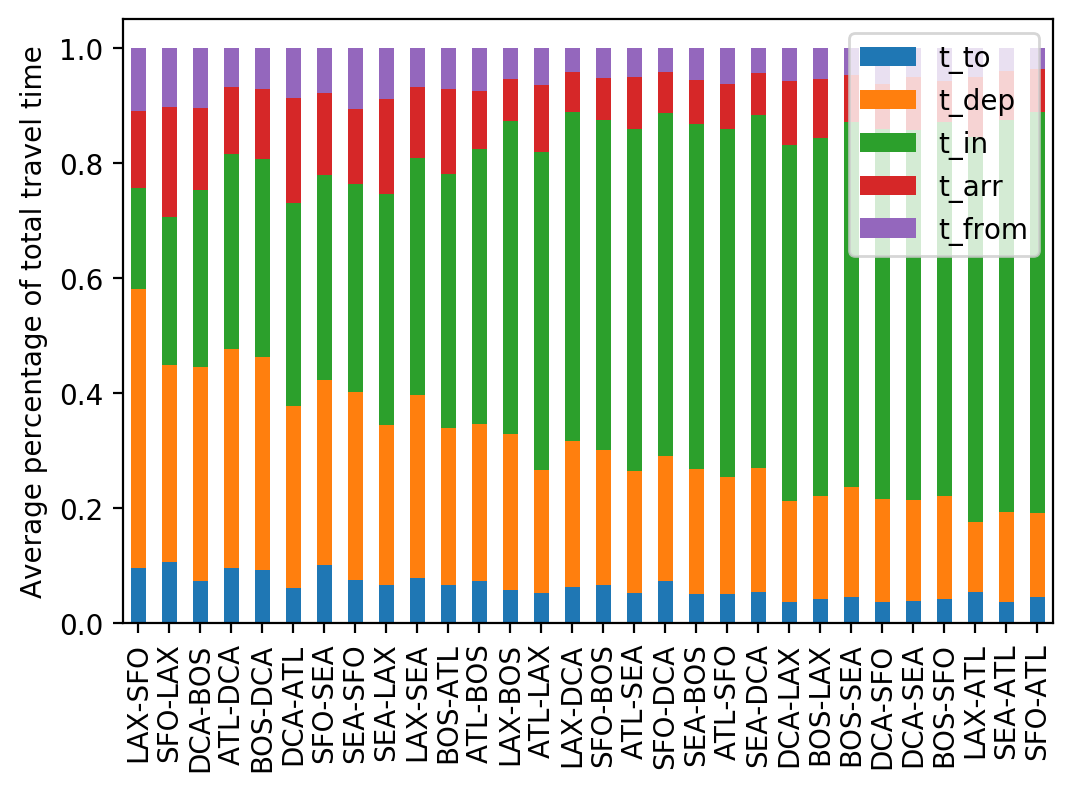}
\caption{Bar plot of the average proportion of the time spent within each phase of the full door-to-door journey for all thirty considered trips.}
\label{fig:bar_time_split}
\end{figure}

We can use the full door-to-door model to better understand the time spent in each leg proportionally to the time spent on the overall trip. For each trip, we calculate the percentage of time spent at each phase based on the full door-to-door travel time. We then calculate the average percentage time spent for each phase and for each city pair trip. Figure\,\ref{fig:bar_time_split} shows the bar plot of these average percentage times for the thirty considered city pairs. The city pairs are sorted according to the percentage of time spent in the actual flight phase. 
With the proposed full door-to-door model, for all considered trips, passengers spend on average more time at the departure airport than riding to and from the airports. This figure also shows that, with this model, for some short-haul flights, such as between SFO and LAX or between BOS and DCA, passengers spend on average more time at the departure airport than in the plane. Refining the full door-to-door model by considering tailored airport processing times $t_\text{sec}$ at departure depending on the city pair and not only on the departure airport could lead to a different conclusion. However, this modification of the model would require more access to passenger data.

\subsection{Airport integration comparison}
\begin{figure}[h!t]
\centering
\includegraphics[width=\columnwidth]{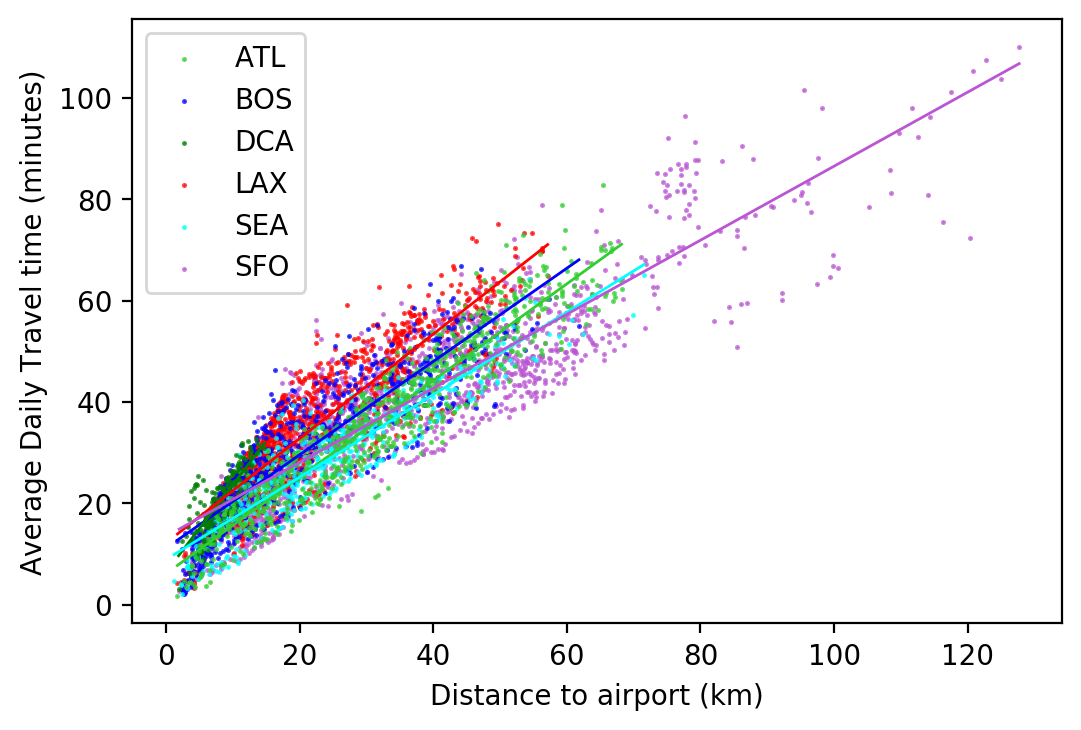}
\caption{Scatter plot of the average ride time to the airport $t_\text{to}$ versus the distance to the airport from January 1\st\, 2018 to March 31\st\, 2018. Straight lines indicate the linear regression fit for each city.}
\label{fig:dist_avg_mean_daily}
\end{figure}

The proposed full door-to-door model allows us to compare each airport's integration within its metropolitan area. Each census tract is associated with an internal point within its boundaries, and this internal point can be used to automatically calculate the distance between airports and each census tract of their metropolitan area. The internal points were defined using an online database\footnote{\url{www.usboundary.com}} based on the US government 2010 census. Figure\,\ref{fig:dist_avg_mean_daily} shows the scatter plot of the average daily ride time to each airport versus the geodesic distance to the airport for the six considered airport. The geodesic distance is the shortest distance between two points along the surface of the Earth. Additionally, the plot also figures a linear regression of these average time with respect to the distance to the airport. A steeper slope for the linear regression indicates that it takes longer to reach the airport from a given distance.

Figure\,\ref{fig:dist_avg_mean_daily} highlights the disparity between the range of each airport within available data: DCA has a range limited to 20 km while SFO attracts Uber riders from more than 120 km away. The other four airports have a similar range. The difference in slope of their associated linear regression is nonetheless useful to rank their integration within their region of attraction. From this perspective, Seattle has the best integrated airport, i.e. the smallest slope, followed by Atlanta, Boston and then Los Angeles.

\subsection{Impact of severe weather analysis}
\begin{figure}[h!t]
\centering
\begin{subfigure}[b]{\columnwidth}
\includegraphics[width=\columnwidth]{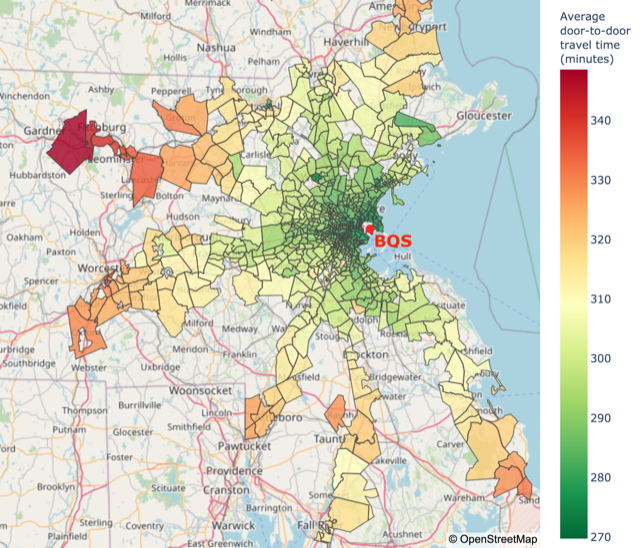}
\caption{Before landfall, on January 2\nd, 2018}
\label{fig:bos_from_dca_02}
\end{subfigure}
\begin{subfigure}[b]{\columnwidth}
\includegraphics[width=\columnwidth]{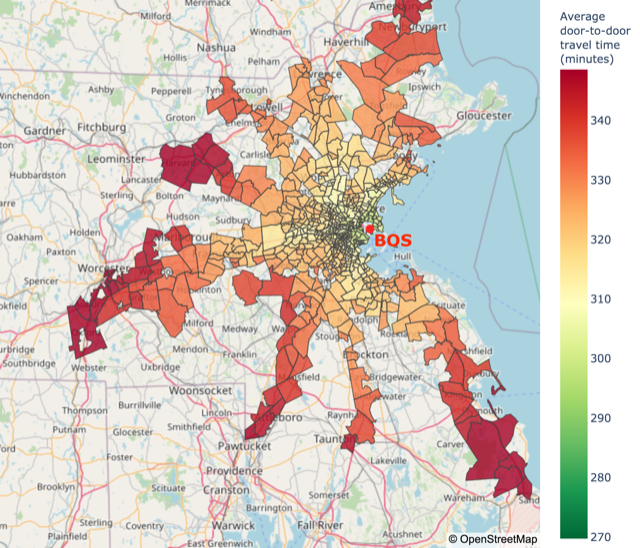}
\caption{After landfall, on January 5\nth, 2018}
\label{fig:bos_from_dca_05}
\end{subfigure}
\caption{Average door-to-door travel times from Washington D.C. city hall to Boston over a single day, before and after the Bomb Cyclone of January 2018.}
\label{fig:bos_from_dca}
\end{figure}

Using the same door-to-door travel time visualization process and applying it to different days can be a tool to better analyze the effects of severe weather perturbations on the full door-to-door journey. As an example, the winter storm previously studied in \cite{marzuoli2018PassengercentricMetricsAir} is analyzed for trips between Washington D.C. and Boston using Figure\,\ref{fig:bos_from_dca}.
This winter storm hit the East Coast of the United States on January 4\nth\, 2018, and led to the closure of two airports in New York City, along with the cancellation of the majority of flights flying to or from the North-Eastern US coast.
Figure\,\ref{fig:bos_from_dca} shows the map of the average full door-to-door travel times to reach the Boston area starting from Washington D.C. city hall on January 2\nd\, 2018, before the landfall of this winter storm, and on January 5\nth\, 2018, after the landfall of the winter storm.

The color scales representing the full door-to-door travel time are identical from one map to another. A shift towards the red is visible from January 2\nd\, 2018 (Figure\,\ref{fig:bos_from_dca_02}) to January 5\nth\, 2018 (Figure\,\ref{fig:bos_from_dca_05}), along with some census tracts disappearing from the considered range on January 5\nth\, 2018, due to lack of sufficient Uber ride data. These two observations indicate that the full door-to-door travel times are closer to the maximum average travel time than from the minimum travel time on January 5\nth\, 2018 compared to January 2\nd\, 2018, and that some zones might have been sufficiently adversely impacted by the weather to prevent rides from the airport to reach them.

\subsection{On the importance of a passenger-centric approach to delays}
A final application to the full door-to-door model presented in this paper emphasizes the difference between flight delay and passenger delay. Since Uber splits the day into five different periods, each with their traffic idiosyncrasies with respect to peak times, we can calculate how much extra travel time is required for a passenger when a flight does not arrive in the scheduled period. For example, a flight expected to arrive in the early morning that lands after 10:00 AM could result in the passenger getting stranded in traffic when trying to leave the airport. Though airlines are not responsible for road traffic, passengers can choose flights based on their arrival time to avoid peak time traffic.

To calculate this extra travel at aggregated level, we have calculated the difference of average travel time between the two periods concerned by flights not arriving according to schedule for each arrival zone. These travel time differences are then aggregated into one travel time difference per city pair by weighing the travel time associated to each census tract with the proportion of passengers initiating their trips from there, or finishing their trip there. The number of passengers originating from or finishing within a census tract is assumed to be proportional to the population density of the considered census tract.

Another measure of sensitivity is to consider the maximum difference between the maximum travel times of each zone between the two considered periods. This second measure indicates the worst variation of the travel time upper bound, i.e. the maximum difference a traveler can experience going from the airport to their final destination zone.

Let us consider the flight UA460 from LAX to SFO scheduled to arrive on Thursday February 15, 2018 at 18:02 local time and that landed with a minor delay of 16 minutes. Due to the 45-minute processing time required to leave the airport, this 16-minute delay shifts the departure period from the airport from afternoon (PM) to late evening. The aggregated average extra travel time is of 15 minutes and 40 seconds, i.e. a 16-minute flight delay triggered an average 31-minute total delay for the passengers. Looking at the second considered measure, the maximum travel time difference for this flight delay is of 72 minutes, meaning that one passenger could experience a delay of 88 minutes resulting from this 16-minute flight delay. This first example illustrates that passenger delays and aircraft delays are distinct.

Paradoxically, arriving earlier than scheduled for a flight does not necessarily mean that the full door-to-door trip ends earlier. For example, flight VX1929 from LAX to SFO scheduled to arrive on Thursday February 8, 2018 at 15:22 local time actually landed 25 minutes earlier. This implied that the passengers were no longer leaving the airport in the afternoon (PM) period but at midday. The aggregated average extra travel time is here of 15 minutes and 2 seconds, so on average travelers did arrive earlier than scheduled, but only by about ten minutes and not the twenty-five minutes announced by the airline. However, looking at the second measurement method again, the maximum ride time difference is 66 minutes and 44 seconds, which means that a passenger could end up arriving forty minutes later than if the flight had landed on time.

\section{Discussion \& Conclusion}
\label{sec:concl}
By leveraging Uber's recently released data and combining it with several other available data sources, this paper proposed a data-driven model for the estimation of full door-to-door travel times for multi-modal trips both in Europe and in the United States. Though the model is used for one city pair in Europe and six different cities in the United States, it can be implemented between any world city pair with sufficient available ride-sharing or taxi data. The proposed model can be adapted depending on how much data about the considered travel modes are available, since a weekly schedule containing no information relative to delays can already lead to some meaningful insights from both a passenger perspective and a city planner perspective.

Once aggregated at a city level, the presented door-to-door travel time model can be used for a paired comparison of the different phases of the full door-to-door journey between two cities. It also enables us to analyze the actual time spent while travelling between two specific cities. The evaluation on a national level of some of the passenger-centric objectives proposed within NextGen in the United States and within ACARE FlightPath 2050 in Europe is possible thanks to the proposed model, especially the objectives regarding how well integrated airports are within their cities. The model can also provide insight to how multi-modal trips are affected by severe weather disruptions, indicating where improvements can be made. It also brings a valuable measurement of the difference between flight delays and passenger delays, emphasizing the need for passenger-centric metrics to evaluate the performance of the air transportation system, which is not solely constituted of planes.

Future studies should consider integrating additional data, such as alternative transportation modes, e.g. the subway, in the model once they are available and when calculating the time needed to reach the departure station or leave from the arrival station. Additionally, the knowledge of the actual daily proportion of passengers travelling via the different approach modes (road or rail) would lead to a better precision of the proposed full door-to-door travel time model. Aggregated information from GPS or mobile phone sources could possibly be used to determine this proportion without infringing passenger privacy.

%

%
\section*{Acknowledgments}
The authors would like to thank Nikunj Oza from NASA-Ames, the BDAI team from Verizon Media in Sunnyvale as well as the \emph{Ecole Nationale de l'Aviation Civile} and King Abdullah University of Science and Technology for their financial support. The authors are also grateful for the help of Marine Lebeater for her feedback. The authors would also like to thank the SESAR Joint Undertaking for its support of the project ER4-10 "TRANSIT: Travel
Information Management for Seamless Intermodal Transport". Data retrieved from Uber Movement, (c) 2020 Uber Technologies, Inc., https://movement.uber.com. 
%

%
\bibliographystyle{IEEEtran}
\IEEEtriggeratref{14}
\bibliography{D2D}%

\end{document}